\documentclass[12pt,a4paper]{article}

\usepackage{epsf,multicol}

\begin{document}

\vspace{1cm}
\begin{center}
{\bf The Influence of Predator-Prey Population Dynamics on the Long-term
Evolution of Food Web Structure}\\

\vspace{0.5cm}

{\it Barbara Drossel$^{1,3}$, Paul G. Higgs$^2$ and Alan J. McKane$^1$} \\
\bigskip
$^1$Department of Theoretical Physics and the $^2$School of Biological 
Sciences,\\ University of Manchester, Manchester M13 9PL, UK \\
$^3$School of Physics and Astronomy, Raymond and Beverly Sackler Faculty 
of Exact Sciences, Tel Aviv University, Tel Aviv 69978, Israel\\
\end{center}


\section*{Abstract}

We develop a set of equations to describe the population dynamics of many 
interacting species in food webs. Predator-prey interactions are non-linear, 
and are based on ratio-dependent functional responses. The equations account 
for competition for resources between members of the same species, and between
members of different species. Predators divide their total hunting/foraging 
effort between the available prey species according to an evolutionarily 
stable strategy (ESS). The ESS foraging behaviour does not correspond to the 
predictions of optimal foraging theory. We use the population dynamics
equations in simulations of the Webworld model of evolving ecosystems. New 
species are added to an existing food web due to speciation events, whilst 
species become extinct due to coevolution and competition. We study the
dynamics of species-diversity in Webworld on a macro-evolutionary timescale.
Coevolutionary interactions are strong enough to cause continuous overturn 
of species, in contrast to our previous Webworld simulations with simpler
population dynamics. Although there are significant fluctuations in species
diversity because of speciation and extinction, very large scale extinction 
avalanches appear to be absent from the dynamics, and we find no evidence for 
self-organised criticality.

\section{Introduction}
\label{intro}

Understanding coevolution within communities of interacting species is
one of the greatest challenges in the study of ecological systems. Two
different sets of issues are involved in modelling such communities. Questions
regarding food web structure, the nature of predator-prey interactions, 
competition for resources, and population dynamics apply on an 
{\it ecological time scale} comparable to the lifetime of individual organisms.
Questions regarding evolutionary change of species, introduction of new 
species to the food web by speciation processes, and removal of species due to
extinction apply on an {\it evolutionary time scale} orders of magnitude 
longer than the lifetime of an organism. We argue that these two sets of 
questions are nevertheless related and that they need to be considered within 
the same framework. In order to understand food web structures we need to 
understand the way in which the diversity of organisms in the ecosystem 
evolved. In order to understand the way the set of species in a food web will 
coevolve we need to understand the nature of the competitive interactions and
predator-prey relationships between them.

The Webworld model, introduced by Caldarelli, Higgs \& McKane
(1998), and studied further here, is an attempt to model the two timescales
simultaneously. The model considers a set of species, each of which has a set 
of morphological and behavioural features that determine the way it interacts 
with all the other species, and hence the positions of links in the food web. 
Population dynamics equations are used to determine the way the population 
sizes of all the species change over ecological time. In one evolutionary time
step of the model one new species is added to the food web, and the 
populations of all the species then change in response. The new species 
sometimes adds stably to the ecosystem, sometimes dies out due to competition 
with existing species, and sometimes causes the extinction of other species. 
The diversity of species within the ecosystem thus changes on the evolutionary
time scale due to speciation and extinction.

In our previous paper (Caldarelli et al (1998), henceforward Paper I), we 
considered the properties of the food webs generated by Webworld, including 
the number of basal, intermediate and top species in the web, the number of 
links per species, and the number of trophic levels. These properties have been
measured in real food webs (e.g. Cohen et al., 1990; Hall \& Raffaelli, 1991;
Goldwasser \& Roughgarden, 1993; Martinez \& Lawton, 1995), and thus an
extensive amount of ecological data exists with which we were able to compare 
the results of the model. The Webworld model generates food webs with
properties that are in reasonable agreement with those of real webs, given the
large uncertainties inherent in measurements on real webs. The model also 
makes predictions about the way food web properties will change as a function
of ecological parameters such as the rate of input of external resources to
the ecosystem, the efficiency of transfer of resources from prey to consumer at
each level of the food chain, and the strength of competition between 
species for the same resources. As such, we feel that the model goes
considerably further than other theoretical models of food web structure,
such as the cascade model (Cohen, 1990; Cohen et al, 1990), which are based on
constructing random graphs. 

The Webworld model also makes predictions about the dynamics of species
diversity that can be compared with the evidence from the fossil record. 
There has been considerable interest in macro-evolutionary models 
recently, generated by the claim that
extinction dynamics are related to the concept of self-organised criticality
(Bak \& Sneppen, 1993; Sol\'e et al. 1996; Sol\'e et al, 1997; Amaral \&
Meyer, 1999). The idea is that the avalanches of extinctions visible in the
fossil record can be expected to arise from the internal coevolutionary 
dynamics of the system, and thus one does not need to postulate catastrophic 
external events such as meteorite strikes or climate changes in order to 
explain the extinctions. In simulations of the Webworld model in Paper I we 
found that large extinction avalanches could occur in situations when the 
ecosystem was poorly adapted to the external conditions, but that as time went
by extinction events became smaller and rarer. The ecosystem tended towards a 
\lq frozen' food web of mutually well-adapted species that could not be 
invaded by new species. These results therefore did not support the idea of 
criticality. 

Whilst models such as that of Bak \& Sneppen (1993) have the merit
of being (deliberately) very simple, we have found that the dynamics of 
ecosystem models depends substantially on the way that such models are set up,
and we feel that it is important to attempt to include some degree of realism 
in the models if one wishes to draw conclusions about the real world. One of 
the aims of this Paper is to develop a general set of equations for population
dynamics that deals with competition between species and predator-prey 
interactions in a food web which can have any arbitrary structure of links 
generated by the evolutionary process. The equations used here are based on 
ratio-dependent functional responses (Arditi \& Ginsburg, 1989; Arditi \& 
Ak\c{c}akaya, 1990; Arditi \& Michalski, 1995) and represent a considerable
improvement on those used in Paper I in the way they treat competition between
predators for a given prey. Another important change is that increased 
adaptation of predators leads to a decrease in prey population size in the 
current model, whereas this was not so in the model used in Paper I. This 
leads to a continuous overturn of new species replacing old ones, in contrast 
to the frozen state found in Paper I. Although the stationary state is now a
dynamical one, we still find no evidence for self-organised criticality.

The outline of the paper is as follows. In section 2 we define
the Webworld model as studied both here and in Paper I. In section 3 we
present the new equations for population dynamics. We show that these
equations satisfy several logical requirements of general food web models.
We also give an interpretation of the choice of diet of predators in terms on 
evolutionary stable strategies. Section 4 gives details of the simulations of 
the Webworld model using the new equations. In sections 5 and 6 we investigate
the long term evolutionary dynamics, firstly in the absence of predators and 
then in webs with predation. We conclude in section 7 with a discussion of the
implications of these changes and with a comparison with other evolutionary 
models.

\section{The Webworld model}
\label{webworld}

This section describes the basis of the Webworld model introduced in I. 
Details that differ from Paper I will be mentioned specifically. Our model is 
a stochastic one, since the characteristics of the species and the speciation 
events are chosen randomly, however we use deterministic dynamical equations 
for the population sizes of each species. A species is defined by the set of 
its morphological or behavioral characteristics. We construct a species in the
model by picking $L$ features out of a pool of $K$ possible features. These 
features represent morphological and behavioural characteristics, which might 
be, for example, \lq\lq nocturnal", \lq\lq having sharp teeth" and 
\lq\lq ability to run fast", however, we do not assign particular biological 
attributes to each feature: they are just integers which run from 1 to $K$. In
our simulations we take $L = 10$ and $K = 500$ for illustrative purposes.

The matrix of scores $m_{\alpha \beta}$, describes how useful one feature, 
$\alpha$, is against any other feature, $\beta$. The $K \times K$ matrix 
$m_{\alpha \beta}$ is antisymmetric (i.e., $m_{\alpha\beta}=-m_{\beta\alpha}$)
and is taken to consist of random Gaussian variables with mean zero and unit 
variance. These are chosen at the beginning of a simulation run and do not 
change during that particular run. The score $S_{ij}$ of one species $i$ 
against another species $j$ is then defined as 
\begin{equation}
S_{ij}=\max\left\{ 0, \frac{1}{L}\sum_{\alpha \in i}\, \sum_{\beta \in j}
m_{\alpha \beta} \right\}\ ,
\label{scores}
\end{equation}
where $\alpha$ runs over all the features of species $i$ and $\beta$ runs over
all the features of species $j$. Thus the score of one species against another 
is essentially just the sum of the scores of the relevant features against each
other. A positive score $S_{ij}$ indicates that species $i$ is adapted for
predation on species $j$, whilst a zero score means that there is no predator-
prey relationship between the species. The scores will be used in the
equations for population dynamics described in the next section. The external 
environment is represented as an additional species 0 which is assigned a set 
of $L$ features randomly at the beginning of a run and which does not change. 
Species having a positive score against the external environment represent 
primary producers of the ecosystem.

The model consists of a changing set of species that may feed on each other 
and on external resources. External resources are input at a constant rate $R$
and are distributed amongst species as a function of their scores, in a way 
that is discussed later. These resources are then tied up in the ecosystem as 
potential \lq\lq food" in the form of prey for predator species. For 
simplicity, we measure resources and population sizes in the same units, so
that $N_i(t)$ denotes the number of individuals for species $i$ at time $t$ or
alternatively the amount of resources invested in species $i$ at this time.  

The short time dynamics is described by a set of equations giving the change 
of population size of any one species in terms of the population sizes of
the other species in the ecosystem. The form of these equations is to be
discussed in the next section. A new species is created by a speciation event 
from one of the existing species. This is carried out by choosing a parent 
species at random, and introducing a new daughter species into the ecosystem 
that differs from the parent species by one randomly chosen feature. The new 
species begins with a population size of 1, and 1 is subtracted from the 
population of the parent species. The populations of all the species are 
determined by iterating the population dynamics equations. If the population 
size of any species falls below one, that species is removed from the system, 
and so rendered extinct. The population dynamics simulation is continued until
all surviving species reach an equilibrium population size, or until a defined
large time period is reached without the populations having reached 
equilibrium. This completes one evolutionary time step of the model, and the 
program proceeds to add another new species. In order to prevent multiple 
copies of identical species from arising, each time a new species is added, a 
check is carried out to ensure that the set of features of the new species is 
not already represented in the ecosystem.   

A minor change between the simulations here and those in Paper I is
the way that species are chosen to undergo speciation: in the original
model they were chosen with a probability proportional to their
population size, here they are chosen randomly. As we will discuss
below, this does not lead to qualitative changes in the behaviour of
the model. The major change in the model is the form chosen for the
population dynamics. This is discussed in detail in the next section.

\section{Population dynamics}
\label{popdyn}

We wish to develop a set of population dynamics equations which is general 
enough to deal with any food web structure. There have been many models of
population dynamics that discuss only two or three species --- e.g. plant plus 
herbivore plus carnivore, or two consumer species competing for the same
resource. However many of these models are not easy to generalise to the 
multiple species case. Most species in an ecosystem are both predators and 
prey and are in competition with several other species. We require equations 
which include all these effects at the same time.

Let the rate at which one individual of species $i$ consumes
individuals of species $j$ be denoted by $g_{ij}(t)$. This is usually
called the `functional response', and it depends in general on the
population sizes. We suppose that the population size of each species
satisfies an equation of the form:
\begin{equation}
\frac{dN_i(t)}{dt} = -N_i(t)+ \lambda \sum_{j}N_i g_{ij}(t) - 
\sum_j N_j g_{ji}(t).
\label{popsize}
\end{equation}
The first term on the right represents a constant rate of death of
individuals in absence of interaction with other species. The final
term is the sum of the rates of predation on species $i$ by all other
species, and the middle term is the rate of increase of species $i$
due to predation on other species. Where there is no predator-prey
relationship between the species the corresponding rate $g_{ij}$ is
zero. For primary producers the middle term includes a non-zero rate
$g_{i0}$ of feeding on the external resources.  The factor $\lambda$
is less than 1, and is known as the ecological efficiency.  It
represents the fraction of the resources of the prey that are
converted into resources of the predator at each stage of the food
chain. Throughout this paper, we have taken $\lambda=0.1$, a value
accepted by many ecologists (Pimm, 1982). We have deliberately chosen the 
form of Eq.~(\ref{popsize}) to be the same for all species.  We do not 
want to define different equations for primary producers, herbivores, 
and carnivores etc, because species can change their position in the
ecosystem as it evolves, and most species are both predators and prey.
For simplicity, we have set the death rate to be equal for all species
and the value of $\lambda$ to be equal for all species. In a more
complex model we could have allowed these quantities to be functions
of the sets of features of each species and then these parameters
would have been subject to evolution in the same way that the
interactions scores between species are subject to evolution.  The
choice of the death rate to be unity in (\ref{popsize}) essentially
sets the time scale for the population dynamics: the time appearing in
this equation has been scaled so that the coefficient of $-N_{i}(t)$
is one.

Equation (\ref{popsize}) is different from Eq.~(5) in Paper I, which
had been designed to be as simple as possible, and to have only one
stationary state. First, Eq.~(5) in Paper I is discrete in time, while
Eq.~(\ref{popsize}) is continuous. However, this difference is of
minor importance, as Eq.~(\ref{popsize}) has to be discretized anyway
for performing computer simulations, giving
Eq.~(\ref{popsize_discrete}) below, which has a similar form to
Eq.~(5) of paper I in the case of large time steps, $\Delta t = 1$.
The second and main difference to Paper I consists in the fact that
the term describing decrease in population size due to predation in
Paper I was chosen to be independent of the rate at which the species is
consumed by its predators. Also, the form of the functional response in
the consumption term had not been chosen according to ecological
considerations.

The main question to be addressed in this section is how to choose a reasonable
function $g_{ij}$ that is applicable for all the situations that arise in the 
ecosystem. Since the final form we use is relatively complex, we will describe 
several particular cases first and build up to the general case.

For a single predator $i$ feeding on a single prey $j$ we suppose
\begin{equation}
g_{ij}(t) = \frac{S_{ij}N_j(t)}{bN_j(t)+S_{ij}N_i(t)}.\label{simple}
\end{equation}
This is known as a ratio-dependent functional response (Arditi \& Ginsburg, 
1989; Arditi \& Michalsi, 1995), because $g_{ij}$ can be written as a function 
of the ratio of prey to predators if both top and bottom are divided by the 
predator population $N_i$. When the prey is very abundant,
$g_{ij}=S_{ij}/b$, i.e. each predator feeds at a constant maximum rate. When 
predators are numerous compared to the available prey, there is competition
between predators, and the rate at which each individual predator can feed
on the prey becomes limited by the amount of prey. In this limit the combined 
rate of consumption of all predators is $N_ig_{ij} = N_j$. This situation is 
known as donor control. Arditi \& Ak\c{c}akaya (1990) have shown that 
interference between predators is significant and that the ratio-dependent 
functional response can be applied to a wide range of real species. 

In our model, the same equation is used to treat consumption of external 
resources by a primary producer. In this case $i$ is the primary producer, and 
the external resources are $j = 0$, with a value of $N_0$ that is kept fixed 
and equal to $R/\lambda$. By writing down the differential equation (2) for the 
case of a single primary producer species 1, we find that the population size 
$N_1$ reaches a stationary value when $\lambda g_{10} = 1$. Hence, the 
equilibrium population size is
\begin{equation}
N_1 = (\lambda S_{10} - b)N_0/S_{10}\ , \label{pop1}
\end{equation}
provided this is positive, i.e. species 1 can only survive if 
$S_{10} > b/\lambda$. Thus $b$ is an important parameter of the model that
determines the minimum score necessary for a consumer to survive on a given 
resource. With the choice of $N_0$ given above, $N_1$ tends to $R$ if it is 
very well adapted ($S_{10} \gg 1$). The parameter $R$ represents the fixed rate of 
supply of non-biological resources, principally sunlight. These resources are 
renewed constantly and hence are never depleted. Also they cannot accumulate 
if not used, hence there is no differential equation for the rate of change
of $N_0$.

If there are additional species competing with $i$ for predation on species 
$j$, equation (\ref{simple}) can be generalised as follows:
\begin{equation}
g_{ij}(t) = \frac{S_{ij}N_j(t)}{bN_j(t)
+\sum_k \alpha_{ki}S_{kj}N_k(t)}.\label{competition}
\end{equation}
The sum in the denominator is over all species $k$ which prey on $j$, including
$i$ itself, i.e. it is over all species for which $S_{kj} > 0$. The
additional predator populations are present in the denominator because
each individual of species $i$ is in competition with the other species as well
as with other members of its own species. The factor $\alpha_{ki}$ is 
introduced to represent the fact that competition between members of the same 
species for a resource is usually stronger than competition between members of 
different species. Thus $\alpha_{ki} < 1$ when $k$ and $i$ are not equal and 
$\alpha_{kk} = 1$ for all $k$. We will in addition suppose that 
$\alpha_{ki} = \alpha_{ik}$. Although addition of this extra factor 
complicates the equations, it is actually essential in order to permit
coexistence of competing species. As an example consider two species 1 and 2 
competing for external resources. In this case:
\begin{equation}
g_{10}(t) = \frac{S_{10}N_0(t)}{bN_0(t) +
S_{10}N_1(t) + \alpha_{12}S_{20}N_2(t)},\label{twoa}
\end{equation}
\begin{equation}
g_{20}(t) = \frac{S_{20}N_0(t)}{bN_0(t) +
S_{20}N_2(t) + \alpha_{12}S_{10}N_1(t)}.\label{twob}
\end{equation}
In the stationary state $\lambda g_{10} = 1$ and $\lambda g_{20} = 1$, hence
\begin{equation}
N_1 = \frac{N_0(\lambda (S_{10}-\alpha_{12} S_{20}) - b(1-\alpha_{12}))}
{S_{10}(1-\alpha_{12}^2)}, \label{twoc}
\end{equation}
\begin{equation}
N_2 = \frac{N_0(\lambda (S_{20}-\alpha_{12} S_{10}) - b(1-\alpha_{12}))}
{S_{20}(1-\alpha_{12}^2)}. \label{twod}
\end{equation}
For coexistence of the two species both the above must be positive, thus the
species can only coexist if
\begin{equation}
-(1-\alpha_{12})(S_{10}-b/\lambda) < S_{20}-S_{10} < 
(1-\alpha_{12})(S_{20}-b/\lambda). \label{twoe}
\end{equation}
Therefore the range of the difference between scores for which coexistence is 
possible is proportional to $1-\alpha_{12}$.
If the competition between species is reduced ($\alpha_{12}$ is reduced) 
it becomes easier for species to coexist on the same resources. If 
$\alpha_{12}=1$ only species with identical scores can coexist. Since in 
general there is more than one predator per prey in real food webs, it is 
necessary to introduce the $\alpha$ parameter into the model. The result that 
species can only coexist if between species competition is weaker than within 
species competition is also found in other models (e.g. Renshaw (1991) pp 
137-139).

For the purpose of our simulations we will suppose that the strength of 
competition depends only on the degree of similarity between the species:
\begin{equation}
\alpha_{ij}=c+(1-c)q_{ij}, \label{overlap}
\end{equation}
where $c$ is a constant such that $0 \le c <1$, and with $q_{ij}$ being the 
fraction of features of species $i$ that are also possessed by species $j$.

We now consider the case of a single predator with more than one prey. It
might seem that we could use equation (\ref{simple}) for each prey $j$.
However this is unsatisfactory, as explained below. In fact we use:
\begin{equation}
g_{ij}(t) = \frac{S_{ij}f_{ij}(t)N_j(t)}{bN_j(t)
+S_{ij}f_{ij}(t)N_i(t)}.\label{gij2}
\end{equation}
where we have introduced the factor $f_{ij}$, which is the fraction of its
effort (or available searching time) that species $i$ puts into preying on
species $j$. These efforts must satisfy $\sum_j f_{ij}=1$ for all $i$.
The importance of introducing the efforts can be understood by considering a 
single predator $i=3$ with two prey $j=1$ and $j=2$ of population sizes $N_1$ 
and $N_2$. In the particular case where the prey are equivalent from the 
predator's point of view (i.e. $S_{31}=S_{32}$), the dynamics of the predator 
population should be identical to the case where there is just one prey 
population of size $N_1+N_2$. This is a `common sense' condition that has been 
emphasised by Arditi \& Michalski (1995) and Berryman {\it et al} (1995), who 
have shown that many dynamical equations used previously did not satisfy the 
condition. In our case, since the prey are equivalent, the predator sets its 
efforts to be proportional to the population sizes: $f_{31}=N_1/(N_1+N_2)$ and
$f_{32}=N_2/(N_1+N_2)$. Calculating the predation rates from 
equation (\ref{gij2}) the total input to the predator can be shown to be
\begin{equation}
g_{31} + g_{32} = \frac{S_{31}(N_1+N_2)}{b(N_1+N_2) + S_{31}N_3},\label{sumtwo}
\end{equation}
which is the same as for a combined species of population $N_1+N_2$. If the
efforts had not been introduced, i.e. equation (\ref{simple}) had been used 
instead of equation (\ref{gij2}), this condition would not be satisfied.

There is an additional reason why it is necessary to introduce the $f_{ij}$. 
The rate of decrease of a prey population $j$ caused by a predator $i$ is
$N_ig_{ij}$. If the simple equation (\ref{simple}) is used, then the effect of 
the predators on the prey is very large when the predator population is large
compared to the prey. When a new species evolves, it always begins from small
population numbers. Usually, there is an existing species which has some
ability to act as a predator on the new species. The existing predator has
a relatively large population because it must already be successfully feeding 
on an established prey species in the ecosystem. Thus the new prey species 
suffers from an enormous level of predation and almost always becomes extinct 
as soon as it is introduced, even if it is substantially better adapted than 
the established  prey. This prevents a diverse ecosystem from evolving. This 
problem is solved by introducing the efforts, since initially the predator 
puts very little effort into feeding on the new prey because there are so few 
of them. The effect of the predator on the new prey is thus in proportion to 
the prey's population size. This permits newly-evolved species to enter the 
ecosystem in a reasonable way.

We now require a rule by which predators assign their efforts to different prey
when the prey are not equivalent. We suppose that the efforts of any species 
$i$ are chosen so that the gain per unit effort $g_{ij}/f_{ij}$ is equal for 
all prey $j$. If this were not true, the predator could increase its energy 
intake by putting more effort into a prey with higher gain per unit effort.
This choice of efforts leads to the condition
\begin{equation}
f_{ij}(t) = \frac{g_{ij}(t)}{\sum_k g_{ik}(t)}.\label{eff}
\end{equation}
It is shown in the appendix that this choice is an evolutionarily stable 
strategy (ESS) (Parker \& Maynard Smith, 1990). If the population has efforts chosen 
in this way, there is no other choice of efforts that can do better, i.e. no 
other strategy than can become more common if it is rare. When prey are 
equivalent, the ESS solution reduces to setting the efforts in proportion to 
the prey population sizes, as above.

Combining all the above considerations we arrive at the following general
form for the functional response that is used in the Webworld simulations in
this paper:
\begin{equation}
g_{ij}(t) = \frac{S_{ij}f_{ij}(t)N_j(t)}{bN_j(t)
+\sum_k \alpha_{ki}S_{kj}f_{kj}(t)N_k(t)},\label{gij}
\end{equation}
with the efforts given by Eq.~(\ref{eff}).

We have previously shown that a restricted form of Eq. (\ref{gij}) was
invariant under aggregation of identical prey. It is straightforward to 
demonstrate that this holds for the general form (\ref{gij}) too. 
The invariance under aggregation of identical predators can also be shown. 
If predator $i$ and predator $l$ are identical, we have 
$S_{ij}=S_{lj}$ and $\alpha_{ij}=\alpha_{lj}$ for all $j$, and $\alpha_{il}=1$.
The combined effect of the two species on prey $j$ is therefore
\begin{displaymath}
N_ig_{ij} + N_lg_{lj}= \frac{S_{ij}N_j(N_if_{ij}+N_lf_{lj})}
{bN_j+S_{ij}(N_if_{ij}+N_lf_{lj})+\sum_{k\neq l,i}\alpha_{ki}f_{kj}S_{kj}N_k},
\end{displaymath}
which is 
obviously identical to the effect
of one predator species of population size $N_i+N_l$ and effort
$$(N_if_{ij}+N_lf_{lj})/(N_i+N_l).$$
Therefore our equations Eq.~(\ref{popsize}) and Eq.~(\ref{gij}) satisfy the
logical requirements of invariance under aggregation of identical species.
We now go on to consider the practicalities of implementing these equations in
simulations.

\section{Implementation of the Webworld program}
\label{implementation}

Solution of time-dependent differential equations involves a numerical 
algorithm such as the Runge-Kutta method which integrates forward by small 
time steps. We require to simulate the dynamics of large numbers of species 
over large times, hence the efficiency of the algorithm is important. To speed 
up the computer simulations, we used a discrete version of the dynamics,
\begin{equation}
N_i(t+\Delta t) = N_i(t)(1-\Delta t) + 
\Delta t\left[\lambda \sum_{j}N_i g_{ij}(t) - \sum_j N_j g_{ji}(t)\right],
\label{popsize_discrete}
\end{equation}
with a time-step $\Delta t = 0.2$, which is quite large. The discrete version 
of the dynamics would only be identical to the differential equation if 
$\Delta t$ were very small. However, for our purposes the continuous and 
discrete time versions provide an equally good description of an ecosystem and 
we do not wish to distinguish between them. A key point about the equations is 
that the stationary values of the population sizes from equation
(\ref{popsize_discrete}) are identical to those from equation (\ref{popsize}).

In the program it is necessary to continuously update the efforts for each
species so that they remain close to the ESS values. We assume that efforts
can change on a time scale of days, which is much quicker than the change in 
population sizes, which occurs on the time scale of the generation time of the 
organism. If the efforts satisfy Eq.~(\ref{eff}),
then they also satisfy the ESS condition that the gain per unit effort is equal
for all prey. If we begin with some choice of efforts that is not the ESS and 
substitute these into the $g_{ij}$ functions on the right of equation
(\ref{eff}) we obtain a new set of efforts that is closer to the ESS. 
Repeated iteration of this equation therefore causes the efforts to converge
on the ESS. In our simulations, after each update of the population sizes using
equation (\ref{popsize_discrete}), we updated the efforts by iterating 
equations (\ref{gij}) and (\ref{eff}) many times whilst keeping the population 
sizes fixed, until the maximum relative change in effort was smaller than a 
threshold. In most of our simulations, this threshold was $10\%$. Only then 
did we proceed with the next update of the population sizes.  

In principle, a species $i$ can assign part of its effort to any
species $j$ for which $S_{ij} > 0$, however, the ESS condition means
that not all such species end up with a non-zero fraction of the
effort. From (\ref{gij}), the gain per unit effort has a limit as
$f_{ij}$ tends to 0, and this is the maximum achievable value of
$g_{ij}/f_{ij}$. If this maximum value is less than the gain per unit
effort that can be achieved from some other combination of prey
excluding $j$, then the ESS solution has $f_{ij}=0$, i.e. $i$ does not
include $j$ in its diet. We believe from numerical studies that there
is a unique ESS choice of efforts for any fixed set of population
sizes. We have proved this in some special cases, but have not yet
found a general proof.

The efforts of each species change continuously during the simulation. If 
$f_{ij}=0$ at some point in time, it does not necessarily remain so. If the 
population size of species $j$ increases it may pay $i$ to switch some of its 
effort to $j$. Also if a third species goes extinct which was a well-adapted 
predator of $j$, it may be possible for $i$ to feed on $j$, whereas previously 
it could not do so because it was outcompeted by the third species. In cases 
where the ESS solution for a particular effort is zero, iteration of equation
(\ref{eff}), from a small starting value, causes it to become ever closer to 
zero, and eventually the computer sets it to zero when it falls below the 
smallest real number allowed by the operating system. This creates a problem, 
since if an effort is exactly zero, it can never increase again by iteration 
of equation (\ref{eff}). Therefore we introduced a minimum effort $f_{\rm min}$
in the simulation program, such that whenever the value of an effort $f_{ij}$ 
became smaller than $f_{\rm min}$, but the score $S_{ij}$ was positive, we set
$f_{ij}=f_{\rm min}$. This allows efforts that were previously effectively zero
to recover again to large values if conditions change. In cases where the
score $S_{ij}$ is zero, then the corresponding effort is also exactly zero.
In these simulations we chose $f_{\rm min}=10^{-6}$, and found that the results 
do not depend on the precise value of $f_{\rm min}$, as long as it is small 
enough.

Beginning from the same ratio-dependent functional response for one predator and
one prey (equation \ref{simple}), Michalski \& Arditi (1995) and Arditi \& Michalski (1995) generalised
the equation to a food web in a way that is different from our equation 
(\ref{gij}). These authors introduced quantities $X_i^{r(j)}$, the part of 
species $i$ that is being accessed as a resource by species $j$, and 
$X_j^{c(i)}$, the part of species $j$ that is acting as a consumer of species 
$i$. Similar quantities were used by Berryman et al. (1995) with a different 
form of the functional response. The values of these quantities must be 
determined self-consistently at each moment in time. This is qualitatively 
similar to the way in which we determine the solution for the efforts $f_{ij}$ 
at each moment in time. Our equations are simpler because they only require 
one set of auxiliary variables per species rather than two. We are also able 
to give an interpretation of the $f_{ij}$ in terms of the ESS, which was not 
done with the alternative formulation using the $X$ parameters. The two 
formulations nevertheless predict similar effects. Michalski \& Arditi (1995) 
show that as the $X$ parameters change, links can appear and disappear from 
the food web, and hence that the structure of the web is not the same at 
equilibrium as away from equilibrium. The same behaviour is seen in our
approach, since the $f_{ij}$ can change from zero to non-zero and vice-versa.
 
\section{Competition for external resources in absence of predation}
\label{nopred} 

We begin by considering the competition between primary producer species for
the external resources $R$ in the absence of consumer species at higher levels
of the food web. We wish to determine how great a degree of diversity can be
generated by evolution in this case, and to study the way the competition
strength affects this diversity. All species feed on the external resources 
(``species 0'') only, therefore $S_{i0} > 0$. All the scores $S_{ij}$ for 
interaction between species are set to zero in this case. Since each species 
has only one food source, all the efforts are identical to 1. We initialize 
the model by assigning the values of the feature score matrix 
$m_{\alpha\beta}$, and by choosing 10 random features to be the features of 
the environment. We then introduce the first species with 10 randomly chosen 
features and a population size $N_1(0)=1$. After iterating the population 
equations until all population sizes converge, a new species is created by 
speciation, as described in section \ref{webworld}, and this process is 
repeated for many evolutionary time steps.

For survival as a primary producer, a species must have a score
$S_{i0} > b/\lambda$, as shown by equation (\ref{pop1}) above. In
addition species can coexist only if their scores are sufficiently
close. The conditions for coexistence in the case of just two species
were given in equation (\ref{twoe}). Species with scores which are too
low are out-competed, and become extinct. The strength of competition
$\alpha_{ij}$ between species depends on their degree of similarity
$q_{ij}$ as defined in equation (\ref{overlap}). Thus a species with a
relatively low score that is phenotypically distant from its
competitors (shares few features) experiences reduced competition and
may survive, whereas another species with the same score that is
similar to a well-adapted high-score species may be out-competed.  The
rationale behind this is that different species can use resources in
different ways. If plants diversify by being of different sizes, by
adapting to different temperatures and moisture levels, and by
adopting different means of dispersal etc., they can make more
effective use of the fixed amount of sunlight and ground space that is
available. When $c$ is close to 1 there is strong competition even
between distantly relates species. When $c$ is small there is much
weaker competition between distant species, hence we would expect
greater diversification of the ecosystem when $c$ is small. 

The results of one simulation run are shown in Figure 1. One can see that the 
species configuration becomes fixed after approximately 13000 time steps. We 
continued the simulation to 100000 time steps and did not see any change. 
Obviously, the species configuration is such that no new species can be 
generated that can survive in the presence of all the existing species. 
However, 
the new species generated are all similar to existing species (they differ by 
only one feature out of 10 from their parent species). Thus the set of species 
that arises is not necessarily stable against all possible species, just 
against those which can arise by small changes of the existing set. If we use 
the same score matrix $m_{\alpha \beta}$ and the same set of features for the 
external environment, but we start with a different initial species, the 
stationary configuration is different. We also found that the surviving 
species with the highest score is usually not the one with the largest 
population. This is because of the dependence of competition on similarity. 
The species with the highest possible score against the environment is usually 
not part of the stationary configuration. For example, when $c=0$, we found
that the mean score of the species in the stationary configuration is below 
6.0, while the best possible score is close to 8.0 for our choice of the 
environment and the score matrix. The similarity between the majority of pairs 
of species was $q_{ij} = 0$. When the simulation was started with the species 
with the highest score, it died out quickly, but the mean stationary score was 
higher than with a random initial species. This situation was different when 
$c$ was larger. For $c\ge 0.8$, the advantage of being different was smaller, 
and the stationary population contained the species with the highest possible 
score, together with other species with a high degree of similarity to it.

These results in absence of predation confirm the importance of the parameter
$c$, and show that reduction in competition as similarity between species 
decreases is an important factor in promoting the evolution of diversity.
They also show that in absence of predation the ecosystem evolves towards
frozen state that cannot be invaded by new species.

\section{Webs with predation} 
\label{fullweb}

In this section we simulate the full Webworld model in the presence of
predation. We start the simulations again with a single primary
producer species, but as the species diversify, webs with several
trophic layers are created. Even though the network of interactions
between species can be complex, we find that the iteration of the
population dynamics equations usually converges rapidly to a fixed
point. This property was also shared by the simpler set we used in 
Paper I. This means that we can wait for convergence of all the population 
sizes for each set of species before the next species is added. More 
complex dynamics is found in other food web models such as 
the one discussed by Blasius {\it et al} (1999).  Our main interest
here is to determine which species survive in the in the long-term. 
Thus the important
features of the dynamics are whether the newly evolved species can
increase in number initially, and whether any other species drop below
the extinction threshold, $N = 1$.

The following figures and data show a selection of simulation
results.  Figure 2 shows the number of species in the system as
function of time, measured again in speciation events. The two curves differ
in the set of random numbers, but not in the parameter values. 
This means that they differ in their score matrix, 
in the features of the external environment and of the first species.

In contrast to the case without predation, the web has a continuous
overturn of species even after a long time. This result is different
from Paper I, where the species configuration became so well adapted
that it could not be invaded by new species. From Fig.~\ref{fig2}, one
can also see that simulations with different random numbers give rise
to webs with similar species numbers and fluctuation strength. It appears that
the differences between runs with the same parameter values are smaller than
they were with the previous equations (c.f. figure 3 in Paper I). 
We also looked at other quantities besides the
species numbers, such as those shown in Table 1 below, and they are
similar for the two simulation runs. 

Next, we studied the influence of the parameter $c$ in
Eq.~(\ref{overlap}) on the properties of the web. As discussed earlier
in this paper, a smaller value of $c$ leads to less competition
between species, and it promotes
diversity. Figure \ref{fig3} shows the number of species as function of time
for four different values of the parameter $c$, all other parameters being
equal. One can see that the species number increases with decreasing $c$, 
due to the decrease in competition. We have argued before that as $c$ 
decreases, the efficiency in exploiting a food source depends more on the 
overlap of a species with its competitors, and less on its score. This is 
demonstrated in Figure \ref{fig4}, which shows selected scores as function 
of time. Initially, the basal species with the highest score $S_{i0}$ is 
chosen and its score is plotted as long as the species exists. When the 
monitored basal species becomes extinct, the basal species with the best 
score at that moment is chosen and monitored, and so on. We have done 
the same for the predator-prey pair with the highest score $S_{ij}$. 
Each step in the curves means that the monitored species has become extinct.
The figure also shows that species overturn is higher on higher trophic 
levels. This is no surprise, since basal species have the largest population 
sizes, and it is therefore more difficult to drive them to extinction.
We find that the scores are higher for larger $c$, and that, in particular,
the basal species are replaced less often for larger $c$. For a value of 
$c = 0.8$ there was an indication that the monitored basal species had 
become fixed.

In Paper I we looked extensively at the structure and statistical
properties of the food webs generated by Webworld, and compared these
to real food webs.  We now wish to look at these properties in the new
simulations to see how the changes in the form of the dynamical
equations influence the web structure.  It turns out that, as in Paper I,
very reasonable agreement with quantities measured in real food webs can be
achieved for certain values of the parameters of the model. For the purpose 
of defining food web structure, we consider a link between species $i$ and 
species $j$ to be present if species $i$ consumes at least one individual 
of species $j$ per unit time, i.e. if $g_{ij}>1$. As in Paper I, we define 
the trophic level of a species to be the number of links on the shortest 
path from the external resources to that species. In the results tables the
Average level is the mean trophic level averaged over all species in the web
and averaged
over time. The Average maximum level is the mean value of the maximum trophic
level in the web averaged over time. In the analysis of food webs, 
species are often classified
according to whether they are basal, intermediate, or top. Basal
species live exclusively on the external resources (i.e. they have no
prey). Top species have no predators.  Intermediate species have both
predators and prey. The following table summarizes the results for
different values of $c$, and for the same parameters as in
Fig.~\ref{fig3}, averaged over several thousand time steps:
\begin{center}
\begin{tabular}{l | r r r r}
$c$ & $0.8$& $0.6$& $0.4$& $0.2$\\
\hline 
No.~of species&27&55&79&196\\
Links per species & 1.68&1.70&2.33&5.33\\
Av.~level & 2.15&2.28&2.38& 2.45\\
Av.~max.~level &4.0&3.91&3.69 & 3.03\\ 
Basal species (\%) & 12& 9& 8 & 2\\
Intermediate species (\%) & 86& 90& 90& 90 \\
Top species (\%) & 2& 1& 2&8\\
Mean overlap level 1& 0.71& 0.37 & 0.22 & 0.06\\ 
Mean overlap level 2& 0.30& 0.13& 0.08 & 0.04\\
Mean overlap level 3& 0.17 & 0.09&0.07& 0.04\\ 
\end{tabular}
\end{center}

\centerline{Table 1. Results of simulations of the model with $R = 10^{5}$
and $b = 5 \times 10^{-3}$}

\centerline{for four values of the competition parameter $c$.}

\medskip

As can be seen from Fig.~\ref{fig3}, the two simulations for $c=0.4$
and $c=0.2$ have not yet reached their stationary state. Nevertheless,
Table 1 shows several trends present: with decreasing $c$, the
fraction of intermediate species, the number of links per species, and
the average trophic level of a species increases, whilst the fraction
of basal species decreases. The mean overlaps on levels 1, 2 and 3 are the 
mean values of the quantity $q_{ij}$ (fraction of shared features) for all pairs
of species on the same level. We observe that the overlap is higher on the lower
levels (i.e. lower level species are more diverse). The mean overlap on each 
level decreases as $c$ decreases, because the strength of competition increases.
The same effect was discussed in the previous section for the case with no
predation.

The effect of the size of the resources on the number of species is shown in 
Figure \ref{fig5}. As would be expected, a larger set of resources can sustain 
a larger web of species. Table 2 shows the mean values of selected properties 
of the web for these simulations:

\begin{center}
\begin{tabular}{l | r r r r}
$R$ &$1.0\times 10^4$ & $1.0\times 10^5$ & $3.5 \times 10^5$& $1.0\times
10^6$\\ 
\hline 
No.~of species&33&57&82&270\\
Links per species & 1.76&1.91&1.91&2.96\\
Av.~level & 1.95& 2.35&2.65& 3.07\\
Av.~max.~level &3.0& 3.9& 4.0& 4.4\\ 
Basal species (\%) & 18 & 9 & 5 &11\\
Intermediate species (\%) & 80& 89 & 89& 89 \\
Top species (\%) & 2&2& 6& 1\\
Mean overlap level 1& 0.32 & 0.34& 0.31& 0.27\\ 
Mean overlap level 2& 0.17 & 0.12& 0.11 & 0.15\\
Mean overlap level 3& 0.19 & 0.09& 0.09& 0.12\\ 
\end{tabular}
\end{center}

\centerline{Table 2. Results of simulations of the model with $c = 0.5$
and $b = 5 \times 10^{-3}$}

\centerline{for four values of the resource $R$.}

\medskip

With increasing size of resources, the number of species, the number of 
levels, and the number of
links per species increase. A larger fraction of species are intermediate
species. An exception is the last simulation (with $10^6$ resources),
which has not yet reached its stationary state.

We also studied the dependence of the model properties on the other
parameters. As $b$ is increased, it is more difficult for a species to
become established, and in particular during the early stages of a
simulation run, the species numbers are smaller. 

When the expression
for the $\alpha_{ij}$ in Eq.~(\ref{overlap}) is modified, the simulation
results are qualitatively similar. We tested explicitly the choice where 
$\alpha_{ij}$ is 1 for $q_{ij}=1$, and a constant smaller than 1 for all other 
$q_{ij}$. 

The figures given in the tables represent averages over several runs. 
The standard deviation
is moderate for links per species and the average level and average maximum
level (about 5\% of the mean), but larger for the number of species and 
the overlaps (about
10\%). The fluctuations in the fraction of basal and intermediate species are
10\% and 5\% respectively, similar to what was found in I. Not
surprisingly, given the small numbers involved, the top species have a large
standard deviation (about 50\% of the mean). For the simulation with the
largest value of $R$, which has not yet fully reached the stationary state,
and where the fluctuations in the number of species are rather large,
the figures given in the table are very rough. In Paper I we made extensive 
comparisons with the statistics of real food webs, hence we will not do that
here. Similar food webs are generated using the equations in the present paper 
to those in paper I.  

As mentioned in Section 2, the rule for speciation can be chosen in
different ways. Instead of choosing each species with the same
probability to be the parent of a new species, we also did a
simulation, where a species was chosen with a probability proportional
to its population size. The mean number of species is smaller if
species undergo speciation in proportion to their population size. The
reason is that more change is happening on the lower trophic levels,
making it more difficult for species in the higher trophic levels to
become established. 

Finally, we studied the size distribution of extinction events. Figure
\ref{fig6} shows the number  $N(s)$  of events for which $s$ species
went extinct during one time step for one long simulation run.
There is a sharp maximum at $s=1$, which is due
to the fact that more than 90\% of the species created by a mutation
cannot survive in the presence of all the other species. The curve has
an exponential decay for larger $s$, indicating that large extinction
events are unlikely. This is very different from the ``self-organised
critical'' behaviour found by Bak and Sneppen (1993) in computer
simulations of a much simpler model for large-scale evolution, where the size
distribution of extinction events follows a power law $N(s)\sim
s^{-\tau}$ with $\tau \simeq 2$.

\section{Conclusion} 
\label{conclusion} 

In this paper, we have studied a model for evolving food webs. We have
established a set of coupled ecological equations for the population
sizes of the different species in a web which satisfies the logical
requirements put forward by Arditi \& Michalski (1995), and in which the 
distribution of foraging effort for each predator follows an evolutionarily 
stable strategy. We have shown that the model generates food web structures
that are comparable to those of real webs, and have considered the trends in 
the web statistics with changing parameter values.

In the absence of predation, the models gives rise to a stable set of
species that cannot be invaded by any close variant species. 
In contrast, in the presence of predation, a web is built that has a 
continuous overturn of species. This result is different from the stable 
species configurations found in Paper I. As the size distribution of
extinction events falls off exponentially, our results are also
different from those of several simpler models for large-scale evolution, which
usually have a size distribution of extinction events that falls of
like a power law with an exponent close to 2 (Bak \& Sneppen, 1993; Amaral
\& Meyer, 1999). 

A central point in the theory of self-organised critical systems is that small
perturbations can sometimes lead to large responses. In the original sandpile
model (Bak {\it et al}, 1988) the addition of one sand grain to the top of a
pile can sometimes lead to avalanches of falling grains. In Webworld, the
equivalent effect is the addition of a new species which can occasionally lead
to several other species becoming extinct. It is important that extinctions
occur as a result of the changes in population sizes caused by adding the new
species. There are no random extinctions in Webworld: we do not remove a
species unless its population falls below 1. There is also no random
replacement of species; when a new species is added, the parent species is not
removed. It may happen that the new species out-competes the parent species and
thus replaces it, but this only happens if the new species is better adapted
than the old one. In contrast, most other macroevolution models (e.g. Bak \&
Sneppen, 1993; Sol\'e et al. 1996; Amaral \& Meyer, 1999) 
either include random extinction or random replacement of species. If
this is done then sooner or later a very well adapted species with a high
population will be removed by chance, and this is likely to have a large effect
on the structure of the ecosystem and maybe lead to further extinctions. In the
sandpile analogy this is like removing a grain from the bottom of the
pile.  It would not be surprising if changes of this type caused large
avalanches. It could be argued that chance extinctions might occur due to
stochastic fluctuations in population sizes. Our dynamics is deterministic and
thus does not allow for this possibility, however stochastic fluctuations are
unlikely to affect species with high population sizes sufficiently to drive
them extinct. Therefore we feel that simply removing species whose populations
fall below the threshold value of 1 is an adequate way of dealing with
extinctions.

One of the major questions that one would wish to address with models such as
ours is whether the large scale extinction events observed in the fossil record
could arise as a result of the internal dynamics of the ecosystem, or whether
external causes are required. We have found in preliminary simulations with our
model (results not shown) that even relatively minor changes
to the external environment (species 0 in our model) are capable of causing 
large scale extinction events, which in turn lead to the potential for the 
growth of new species. Thus it seems clear to us that external perturbations
can cause extinction avalanches. The more interesting question is therefore
whether mass extinctions occur with a static external environment - the case 
considered in this paper. Although the long-term behaviour of our model shows a continual
overturn of species, no large scale extinctions are seen and we therefore
deduce that environmental changes are required to produce these. As we have
stressed above, we believe that what is referred to as ``internal dynamics" in
some models is effectively external, since in these models perfectly well adapted species are 
removed by random changes. So in our view these effects are indistinguishable 
from the elimination of species due to some random external perturbation due 
to environmental change. We conclude that great care has to be taken to 
distinguish between truly internal dynamics and external influences.
For this a realistic model of evolutionary dynamics is required. It is 
likely that both internal and external effects exist in the spectrum of extinction events 
seen in the fossil record.

Another question of interest concerns the robustness of the simulation 
results to modifications of the model. As mentioned at several
places in this paper, we found that our qualitative results are
insensitive to a variety of changes that we made. However, we should
also mention that our findings depend sensitively on a good
implementation of the rule for updating the efforts. 
If the efforts are not given enough
time to equilibrate with respect to the population sizes, small values
of the efforts cannot recover quickly when a
prey becomes more abundant, and we found that this occasionally led to
large extinction
avalanches which destroyed almost the entire web. Another type of undesirable
behaviour was also observed in some simulations where we did not allow
the efforts to equilibrate properly. If the efforts of new
species are initialized such that they are not close to their
equilibrium value, the species configuration becomes frozen after some
time, because no new species can become established. 

There are many other questions to ask: experimental ones which relate to the
comparison with real systems and theoretical ones which have to do with model
structure. We hope to investigate such questions in the future. However, we
believe that the model introduced in this paper is an important step in our
understanding of the evolution of food web structure, being both simple enough 
to give an understanding of the basic mechanisms at work and realistic enough 
to allow comparison with data collected by ecologists.

\section*{Acknowledgements}
This work was supported in part by EPSRC grant K/79307 (BD and AJM) and by the
Minerva Foundation (BD).

\section*{Appendix. Evolutionarily Stable Strategies}

Here we consider a predator species $i$ and we determine the ESS choice of
efforts. Let the total population be $N_i$, and suppose that the majority of
the population, $N_i-n_i$, have a foraging strategy defined by the efforts
$f_{ij}$, whilst a small minority, $n_i$, have a different strategy $h_{ij}$.
Following the usual argument of evolutionary game theory, we require to 
calculate the payoff to the minority and majority strategies, and hence to
determine conditions under which the minority can invade. In this case the 
payoff is the total rate of gain of resources from all prey. The payoff 
for the strategy $f_{ij}$ in absence of the minority strategy is
\begin{equation}
G = \sum_j \frac{S_{ij}f_{ij}N_j}{S_{ij}f_{ij}N_i + K_{ij}} ,\label{appen1}
\end{equation}
where, for convenience, we use $K_{ij}$ to denote all the terms in the
denominator of the $g_{ij}$ function in equation (\ref{gij}) that do not depend
on the efforts of species $i$:
\begin{equation}
K_{ij} = bN_j +\sum_{k \neq i} \alpha_{ki}S_{kj}f_{kj}N_k .\label{appen2}
\end{equation}
The payoff for the minority species in the presence of the majority is:
\begin{equation}
G_{min} = \sum_j \frac{S_{ij}h_{ij}N_j}
{S_{ij}f_{ij}(N_i-n_i) + S_{ij}h_{ij}n_i + K_{ij}} .\label{appen3}
\end{equation}
In the above equation, since the two strategies are played by different
individuals which are members of the same species, the $\alpha$ factor for 
competition between individuals with different strategies is 1, as it is for 
individuals using the same strategy. We require the payoff to the invading 
strategy when it is rare (i.e. when $n_i \ll N_i$), which is just obtained by 
setting $n_i$ equal to zero in (\ref{appen3}). In a similar way the payoff to 
the majority strategy in the presence of the minority can be written down, but 
this reduces to (\ref{appen1}) when $n_i \ll N_i$. This gives
\begin{equation}
G_{min} - G = \sum_j \frac{S_{ij}N_j(h_{ij}-f_{ij})}
{S_{ij}f_{ij}N_i + K_{ij}}
= \sum_j (h_{ij}-f_{ij})\frac{g_{ij}}{f_{ij}} ,\label{appen4}
\end{equation}
where $g_{ij}/f_{ij}$ in the equation above is the gain per unit effort from 
prey $j$ for the majority strategy. The minority strategy can invade if 
$G_{min}-G > 0$. 

Now suppose that the invading strategy differs from the majority strategy for 
two prey species $k$ and $l$, so that 
$h_{ik}=f_{ik}+\Delta f$, $h_{il}=f_{il}-\Delta f$, and $h_{ij}=f_{ij}$ for
all the other prey $j$. In this case
\begin{equation}
G_{min} - G = \Delta f\left(\frac{g_{ik}}{f_{ik}}-
\frac{g_{il}}{f_{il}}\right) .
\label{appen5}
\end{equation}
If the gain per unit effort from prey $k$ is higher than that from prey $l$ 
then any strategy with positive $\Delta f$ can invade, whilst if the reverse 
is true then any strategy with negative $\Delta f$ can invade. However, if the 
gain per unit effort is equal for the two prey then variant strategies are 
neutral. The same can be said for any pair of prey species $k$ and $l$. It 
therefore follows that the ESS is the strategy with the gain per unit effort 
being equal for all prey. If neutral variant strategies accumulate to a 
non-negligible fraction, then selection will again operate to drive the 
population back to the ESS.

It is interesting to note that the ESS does not correspond to the solution 
predicted by optimal foraging theory (OFT) (Stephens \& Krebs, 1986). The
OFT solution would be to maximise $G$ in equation (\ref{appen1}) with the 
constraint that the efforts sum to 1. This can be calculated, and the result 
is (by definition) greater than the total gain to a predator when all adopt 
the ESS. However, a single ESS predator in a population of OFT predators 
actually has a higher total gain than the OFT population. Thus the ESS can 
invade the OFT solution, but the reverse is not true. Hence we argue that the 
ESS is the appropriate choice of efforts for our model. 

In most of the models considered by Stephens \& Krebs (1986) the
payoff to the predator is not affected by what other predators do,
therefore the straightforward OFT solution of optimising the total
rate of energy intake is appropriate (see their comments on p 211
regarding game theory). However, competition between predators of the
same species and between different species is an essential part of the
way our population dynamics equations are set up, and we also believe
it is an important factor in real ecosystems.  Therefore it is
important to treat the foraging problem from a game theory point of
view. The need for game theory has also been defended recently by
Reeve \& Dugatkin (1998). Various game theory models dealing with
aspects of foraging behaviour have been proposed (Matsuda et al.,
1996; Shaw et al., 1995; Leonardsson \& Johansson, 1997; Visser, 1991;
Giraldeau \& Livoreil, 1998; Sih, 1998).

\section*{References}

\noindent Amaral, L.A.N. \& Meyer, M. (1999). Environmental changes, 
co-extinction, and patterns in the fossil record. {Phys. Rev. Lett.} 
{\bf 82}, 652-655.

\noindent Arditi, R. \& Ginzburg, L.R. (1989). Coupling in
predator-prey dynamics: Ratio-dependence. {\it J. Theor. Biol.} {\bf
139}, 311-326.

\noindent Arditi, R. \& Ak\c{c}akaya, H.R. (1990). Underestimation of mutual
interference of predators. {\it Oecologia} {\bf 83}, 358-361.

\noindent Arditi, R. \& Michalski, J. (1995). Nonlinear food web models
and their responses to increased basal productivity. In {\it Food
webs: integration of patterns and dynamics} (ed. G.A. Polis \&
K.O. Winemiller), pp.~122-133, Chapman \& Hall, London.

\noindent Bak, P., Tang, C. and Wiesenfeld, K. (1988). Self-organized 
criticality, Phys. Rev. A{\bf 38}, 364-374.

\noindent Bak, P. \& Sneppen, K. (1993). Punctuated equilibrium and 
criticality in a simple model of evolution. {\it Phys. Rev. Lett.} {\bf 71}, 
4083-4086.

\noindent Berryman, A.A., Michalski, J., Gutierrez, A.P., Arditi, R. 
(1995). Logistic theory of food web dynamics. {\it Ecology} {\bf 76},
 336-343.

\noindent Blasius, B., Huppert, A. \& Stone, L. (1999). Complex dynamics and 
phase synchronization in spatially extended ecological systems. {\it Nature} 
{\bf 399}, 354-359.

\noindent Caldarelli, G., Higgs, P.G., McKane, A.J. (1998). Modelling 
coevolution in multispecies communities. {\it J. Theor. Biol.} {\bf
193}, 345-358.

\noindent Cohen, J.E. (1990). A stochastic theory of community food webs VI -
Heterogeneous alternatives to the Cascade model. {\it Theor. Pop. Biol.} 
{\bf 37}, 55-90.

\noindent Cohen, J.E., Briand, F. \& Newman, C.M. (1990). {\it Biomathematics}
Vol. 20. Community Food Webs, Data and Theory. Springer Verlag, Berlin.

\noindent Giraldeau, L.A. \& Livoreil, B. (1998). Game theory and social 
foraging. {\it Game Theory and Animal Behaviour} 16-37. Eds. Dugatkin, L.A \&
Reeve, H.K. Oxford University Press.

\noindent Goldwasser, L. \& Roughgarden, J. (1993). Construction and analysis 
of a large Caribbean food web. {\it Ecology} {\bf 74}, 1216-1233.

\noindent Hall, S.J. \& Raffaelli, D. (1991). Food web patterns: lessons from
a species-rich web. {\it J. Anim. Ecol.} {\bf 60}, 823-842.

\noindent Leonardsson, K. \& Johansson, F. (1997). Optimum search speed and
activity: a dynamic game in a three-link trophic system. {\it J. Evol. Biol.}
{\bf 10}, 703-729.

\noindent Martinez, N.D. \& Lawton, J.H. (1995). Scale and food web structure -
from local to global. {\it Oikos} {\bf 73}, 148-154.

\noindent Matsuda, H., Hori, M. \& Abrams, P.A. (1996). Effects of 
predator-specific defence on biodiversity and community complexity in 
two-trophic-level communities. {\it Evolutionary Ecology} {\bf 10}, 13-28.

\noindent Parker, G.A. \& Maynard Smith, J. (1990). Optimality theory in
evolutionary biology. {\it Nature}. {\bf 348}, 27-33.

\noindent Michalski, J. \& Arditi, R. (1995). Food web structure at equilibrium
and far from it: is it the same? {\it Proc. R. Soc. Lond. B} {\bf 259},
217-222. 

\noindent Pimm, S. L. (1982). {\it Food Webs}. Chapman \& Hall, London.

\noindent Reeve, H.K \& Dugatkin, L.A. (1998). Why we need evolutionary game
theory. {\it Game Theory and Animal Behaviour} 304-311. Eds. Dugatkin, L.A \&
Reeve, H.K. Oxford University Press.

\noindent Renshaw, E. (1991). {\it Modelling Biological Populations in Space
and Time.} Cambridge studies in mathematical biology, 11. Cambridge University
Press.

\noindent Shaw, J.J., Tregenza, T., Parker, G.A. \& Harvey, I.F. (1995).
Evolutionarily stable foraging speeds in feeding scrambles: a model and an
experimental test. {\it Proc. Roy. Soc. Lond.} B {\bf 260}, 273-277.

\noindent Sih, A. (1998). Game theory and predator-prey response races.
{\it Game Theory and Animal Behaviour} 221-238. Eds. Dugatkin, L.A \&
Reeve, H.K. Oxford University Press.

\noindent Sol\'e , R.V., Bascompte, J. \& Manrubia, S.C. (1996). 
Extinction: good genes or weak chaos? {\it Proc. Roy. Soc. Lond. B}
{\bf 263} 1407-1413.

\noindent Sol\'e, R.V., Manrubia, S.C., Benton, M. \& Bak, P. (1997). 
Self-similarity of extinction statistics in the fossil record. {\it Nature} 
{\bf 388}, 764-767.

\noindent Stephens, D.W. \& Krebs, J.R. (1986). {\it Foraging Theory}. 
Princeton University Press, New Jersey.

\noindent Visser, M.E. (1991). Prey selection by predators depleting a patch:
an ESS model. {\it Netherlands J. Zool.} {\bf 41}, 63-80.

\newpage

\section*{Figures}

\bigskip

\begin{figure}
\centerline{\epsfysize=0.54\columnwidth{\epsfbox{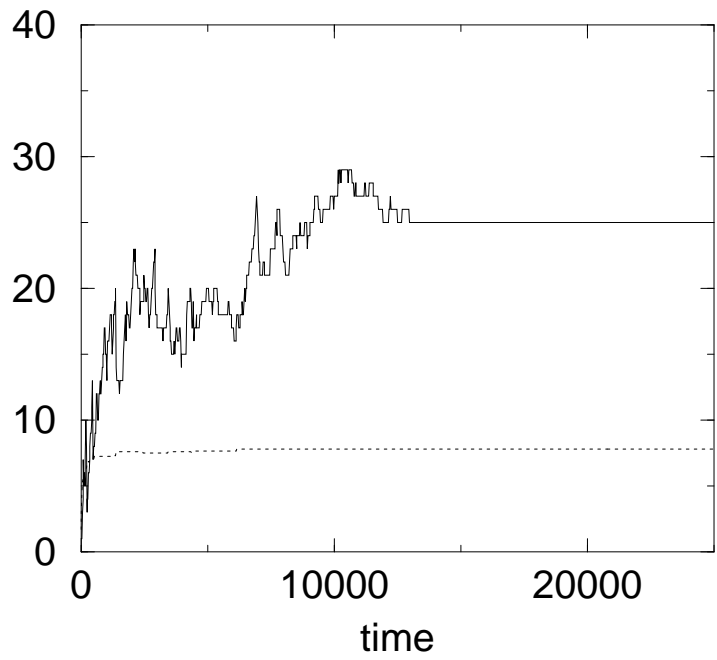}}}
{\caption{Number of species (solid line) and  maximum score (dotted), 
as function of time for $R=1000$, $b=0.1$, 
$c=0.5$. Time is measured in units of number of speciation events.}}
\end{figure}
\begin{figure}
\centerline{\epsfysize=0.54\columnwidth{\epsfbox{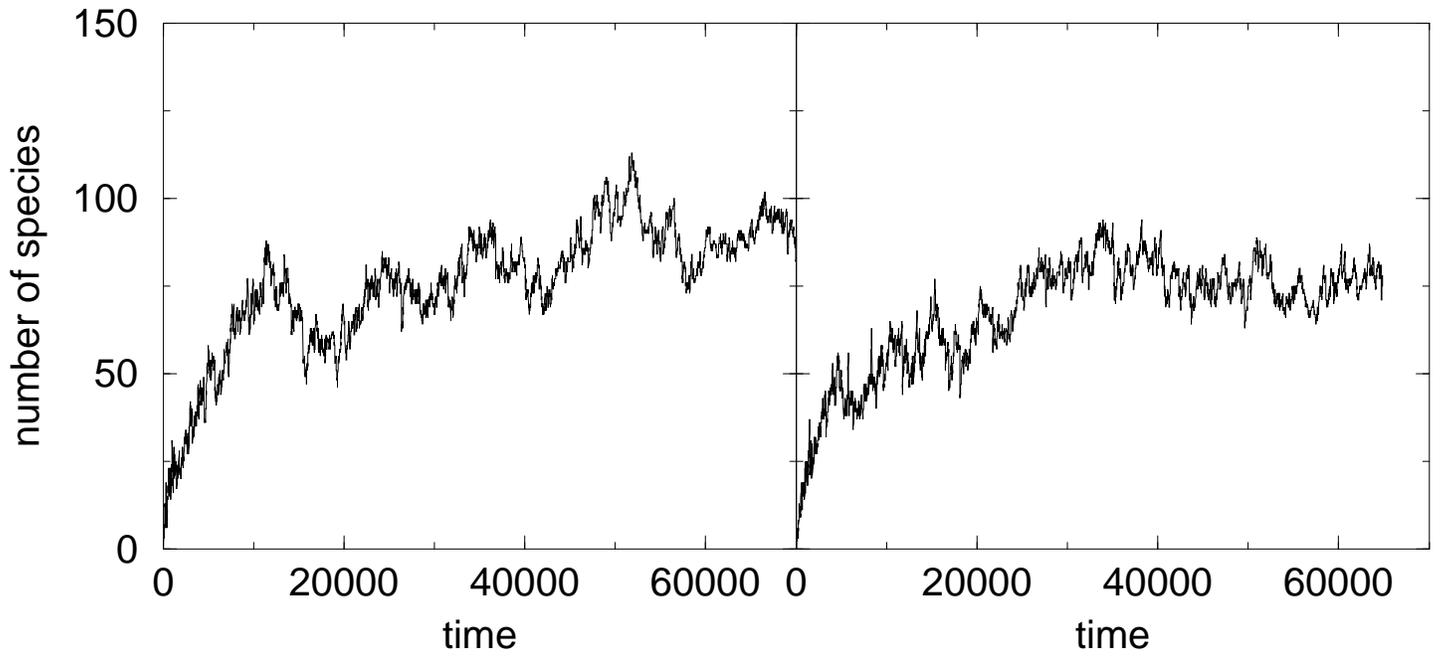}}}
\caption{Number of species as function of time for $R=350000$, $b=0.005$, 
  $c=0.5$. Time is measured in units of mutations. The two simulations
  were carried out with different sets of random numbers. }\label{fig2}
\end{figure}
\begin{figure}
\centerline{\epsfysize=0.96\columnwidth{\epsfbox{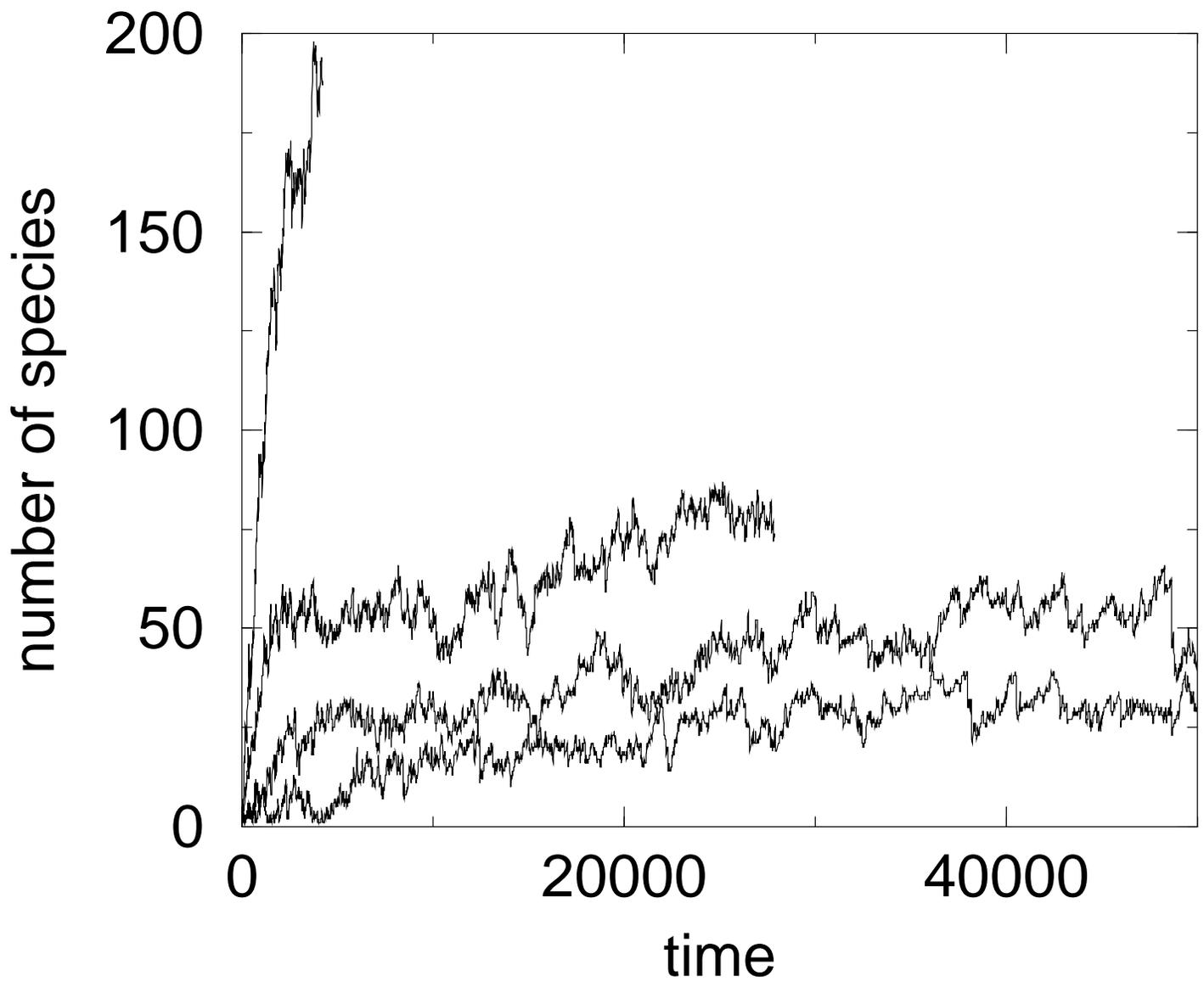}}}
\caption{Number of species as function of time for $R=100000$, $b=0.005$, 
  and $c=0.2, 0.4, 0.6, 0.8$ (from top to bottom curve). Time is
  measured in units of mutations.}\label{fig3}
\end{figure}
\begin{figure}
\centerline{\epsfysize=0.54\columnwidth{\epsfbox{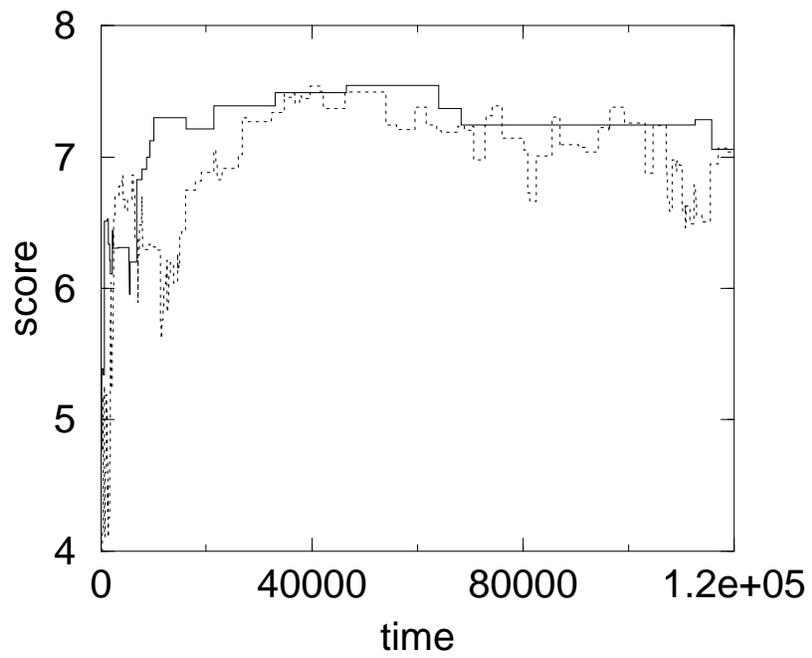}}}
{\caption{Score of the monitored basal species (solid line) and
predator-prey pair (dotted line) as function of time. The parameters 
are $R=100000$, $b=0.005$, and $c=0.5$.}\label{fig4}}
\end{figure}
\begin{figure}
\centerline{\epsfysize=0.54\columnwidth{\epsfbox{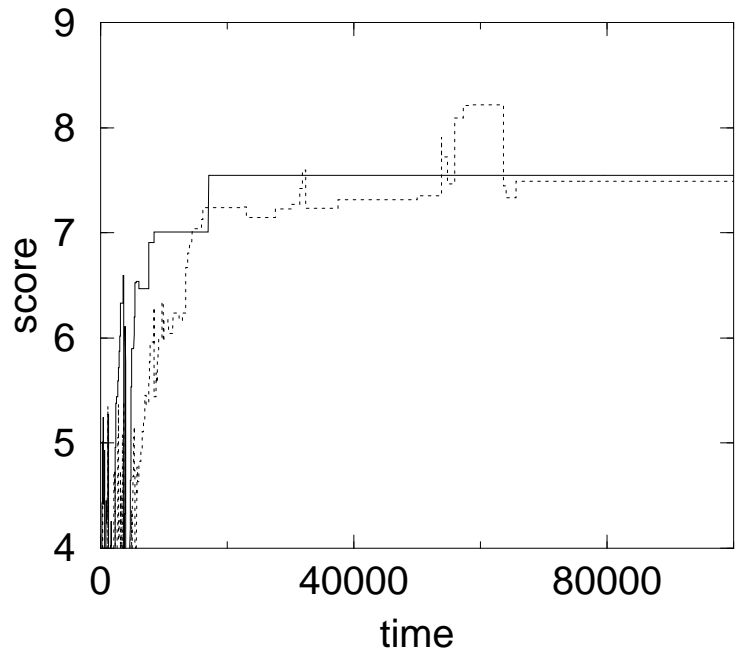}}}
\caption{Number of species as function of time for $c=0.5$, $b=0.005$, 
  and $R=10^4, 10^5, 3.5\times 10^5, 10^6$ (from bottom to top curve). Time is
  measured in units of mutations.}\label{fig5}
\end{figure}
\begin{figure}
\centerline{\epsfysize=0.54\columnwidth{\epsfbox{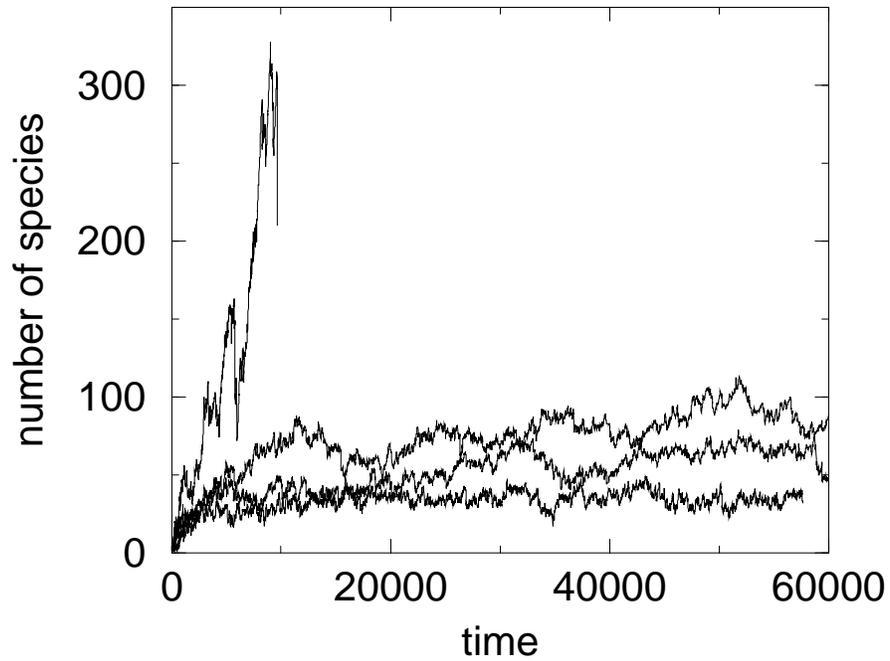}}}
\caption{Size distribution of extinction events for a simulation with
$R=100000$, $c=0.5$, $b=0.005$. Species were chosen with a 
probability proportional to their population size to be the parent of a new 
species.}\label{fig6}
\end{figure}
\end{document}